\documentclass[10pt,aps,prc,twocolumn,floatfix,nofootinbib]{revtex4-1}
\usepackage{amsmath,amssymb,mathrsfs}
\usepackage{bm}
\usepackage{ascmac}
\usepackage{geometry}
\usepackage{graphicx}
\usepackage{color}

\begin{document}


\title{Deep-inelastic multinucleon transfer processes in the $^{16}$O+$^{27}$Al reaction}

\author{B.~J.~Roy$^1$}\email{bjroy@barc.gov.in, bidyutr2003@gmail.com}
\author{Y.~Sawant}
\author{P.~Patwari}
\author{S.~Santra$^1$}
\author{A.~Pal$^1$}
\author{A.~Kundu$^1$}
\author{D.~Chattopadhyay$^1$}
\author{V.~Jha}
\author{S.~K.~Pandit$^1$}
\author{V.~V.~Parkar$^1$}
\author{K.~Ramachandran}
\author{K.~Mahata$^1$}
\author{B.~K.~Nayak$^1$}
\author{A.~Saxena$^1$}
\author{S.~Kailas$^1$}

\affiliation{Nuclear Physics Division, Bhabha Atomic Research Centre, Mumbai 400 085, India}
\affiliation{$^1$Homi Bhabha National Institute, Anushakti Nagar, Mumbai 400 094, India}

\author{T.~N.~Nag}

\affiliation{Radio Chemistry Division, Bhabha Atomic Research Centre, Mumbai 400 085, India}

\author{R.~N.~Sahoo}
\author{P.~P.~Singh}

\affiliation{Department of Physics, Indian Institute of Technology Ropar, Rupnagar, Punjab 140 001, India}

\author{K.~Sekizawa}
\email[Present address: Center for Transdisciplinary Research, Institute for Research Promotion, Niigata University, Niigata 950-2181, Japan; Electronic address: ]{sekizawa@phys.sc.niigata-u.ac.jp}
\affiliation{Faculty of Physics, Warsaw University of Technology, Ulica Koszykowa 75, 00-662 Warsaw, Poland}

\date{January 26, 2018}

\begin{abstract}
The reaction mechanism of deep-inelastic multinucleon transfer processes in the
$^{16}$O+$^{27}$Al reaction at an incident $^{16}$O energy ($E_{\rm lab}=134$~MeV)
substantially above the Coulomb barrier has been studied both experimentally and theoretically.
Elastic-scattering angular distribution, total kinetic energy loss spectra and angular distributions
for various transfer channels have been measured. The $Q$-value- and angle-integrated isotope
production cross sections have been deduced. To obtain deeper insight into the underlying reaction
mechanism, we have carried out a detailed analysis based on the time-dependent Hartree-Fock
(TDHF) theory. A recently developed method, TDHF+GEMINI, has been applied to evaluate production
cross sections for secondary products. From a comparison between the experimental and theoretical
cross sections, we find that the theory qualitatively reproduces the experimental data. Significant
effects of secondary light-particle emissions are demonstrated. Possible interplay between fusion-fission,
deep-inelastic, multinucleon transfer and particle evaporation processes are discussed.
\end{abstract}


\maketitle

\section{INTRODUCTION}\label{sec:intro}

The mechanism of deep-inelastic processes which arise in an intermediate regime
between the direct and the compound-nucleus reactions is notably complicated and has
been a subject of extensive studies for a long time both experimentally and theoretically
\cite{Bass_Book(1980)}. While it offers a precious opportunity to seek for dissipation and
equilibration mechanisms in complex nuclear many-body systems, clear-cut distinction
between deep-inelastic and fusion-fission processes is particularly difficult to draw.
The complexity even increases for light systems, due to a strong overlap between
mass distributions of the fragments from different origins. In this paper, we report
on reaction mechanism studies of deep-inelastic collisions in one of such light systems,
the $^{16}$O+$^{27}$Al reaction.

The quest for understanding of the reaction mechanism with the $^{16}$O+$^{27}$Al
reaction already started as early as 1959 \cite{Anderson(1959)}. Since then, along with
continuous development of experimental techniques, numerous measurements were
carried out for this system to identify, \textit{e.g.}, interplay between incomplete fusion
and projectile breakup processes \cite{Harris(1977),Tsang(1981),Sasagase(1983),
Padalino(1985),Padalino(1986),CarlinFilho(1989),Padalino(1990),Added(1992)}, and
the mechanism of complete fusion \cite{Kowalski(1968),Dauk(1975),Kozub(1975),
Back(1977),Eisen(1977),Rascher(1979),Lee(1981),Ikezoe(1985)}. Measurements for
deep-inelastic and transfer processes were also performed \cite{Ladenbauer-Bellis(1962),
Gridnev(1970),Puhlhofer(1972),Cormier(1976),Mikumo(1980),Shapira(1980),Tabor(1981),
Young(1982),Sato(1983),Brandan(1986),Shen(1987),Bhattacharya(2002),Bhattacharya(2004),RKC(1994)}.
Recently, the elastic scattering of $^{16}$O+$^{27}$Al at 100~MeV was also studied in
the context of nuclear rainbow detection \cite{Cavallaro(2011),Pereira(2012),Oliveira(2013),
Cappuzzello(2016)} that was predicted theoretically in Ref.~\cite{Pereira(2009)}.

In detailed studies of the $^{16}$O+$^{27}$Al reaction at $E_{\rm lab}=88$~MeV
\cite{Mikumo(1980)} and 90, 100~MeV \cite{Cormier(1976)}, it was observed that,
although the system is very light, the overall features of the reaction products are very
similar to much heavier projectile-target combinations at higher collision energies
\cite{Galin(1970),Artukh(1973), Albrecht(1975)}. It indicates that deep-inelastic
transfer processes dominate the reaction and the reaction mechanism evolves gradually
from quasi-elastic to deep-inelastic and to complete fusion. In going from $E_{\rm lab}=90$ to 100~MeV,
it was observed that the deep-inelastic cross section increases significantly. In another
work, the authors of Ref.~\cite{Puhlhofer(1972)} concluded, based on the study of
energy dependence of gamma-ray yields, that up to at least $E_{\rm lab}=165$~MeV
the deep-inelastic transfer reactions will be the primary mechanism and any contribution
from the fusion-fission is probably a small fraction of the total reaction cross section,
though it was an indirect evidence. Contrary to this, a later study of the same system
at $E_{\rm lab}=116$~MeV showed different results \cite{Bhattacharya(2002)}.
In that work, from the study of fragment emission spectra and using a simple two-Gaussian
fitting procedure, the authors deduced deep-inelastic and fusion-fission components separately.
It was observed that the fusion-fission is rather competitive with the deep-inelastic process.
They got the same conclusion from the analysis applied to a neighboring system,
$^{16}$O+$^{28}$Si \cite{Bhattacharya(2004)}.

Despite the considerable progress in this field, the reaction mechanism of deep-inelastic
multinucleon transfer processes is not fully understood. A concept of deep-inelastic transfer
reactions was conjectured in 1970 by Gridnev, Volkov, and Wilczy\'nski \cite{Gridnev(1970)}
in terms of an orbiting picture of a dinuclear complex, where the surfaces of two colliding nuclei
have considerable overlap forming a neck structure during the collision, and it rotates as a whole
and evolves in time. The dinuclear system evolves toward equilibrium by exchanging nucleons
and changing its shape, but it eventually splits into two fragments before the compound-nucleus
formation. Such a classical picture of the dinuclear system formation was successful in explaining
the overall features of experimental angular distributions and isotope production cross sections in
various systems \cite{Volkov(1978)}. On the other hand, in such deep-inelastic processes, reaction
products must be highly excited that leads to secondary particle evaporation. These processes---transfer
followed by evaporation---may contribute significantly to the final yields and have to be carefully
taken into consideration. We should also note that the deep-inelastic heavy ion transfer reactions,
apart from its importance for spectroscopic studies of yrast excitations in nuclei that are not accessible
in standard fusion evaporation processes employing stable nuclei \cite{Broda(2004)}, are recently
considered to be an efficient way to produce and study neutron-rich heavy nuclei in the region of
astrophysical interest \cite{Kozulin(2012), Barrett(2015),Watanabe(2015)}.

In view of its importance and with a motivation to further understand the mechanism of
the deep-inelastic multinucleon transfer reaction, we have carried out an experiment for
the $^{16}$O+$^{27}$Al reaction at $E_{\rm lab}=134$~MeV, an incident energy
significantly above the Coulomb barrier ($E/V_{\rm B}\simeq5.3$, where $V_{\rm B}$
is the phenomenological fusion barrier \cite{Bass(1974),Bass(1977)}). Elastic-scattering
angular distribution, total kinetic energy loss (TKEL) spectra and angular distributions for
various transfer channels have been measured. Employing a microscopic framework of the
time-dependent Hartree-Fock (TDHF) theory and its recent extension, TDHF+GEMINI
\cite{Sekizawa(2017)}, we have theoretically examined the underlying reaction mechanism.
The latter approach combines TDHF with a statistical de-excitation model, {\footnotesize
GEMINI}++ \cite{GEMINI++}, allowing the evaluation of production cross sections for
secondary reaction products. From the present experimental data and detailed analysis
based on the TDHF theory, we aim to elucidate the underlying reaction mechanism in
the $^{16}$O+$^{27}$Al reaction.

The paper is organized as follows.
In Sec.~\ref{sec:expt}, we describe details of the measurement and present the experimental data.
In Sec.~\ref{sec:TDHF}, the results of the TDHF calculations are given and the reaction dynamics is investigated.
In Sec.~\ref{sec:comparison}, total isotope production cross sections obtained from the experiment and those from TDHF(+GEMINI) are compared.
We summarize this work in Sec.~\ref{sec:summary}.

\section{EXPERIMENTAL DETAILS AND RESULTS}\label{sec:expt}

\begin{figure}[t]
\begin{center}
\includegraphics[width=\columnwidth]{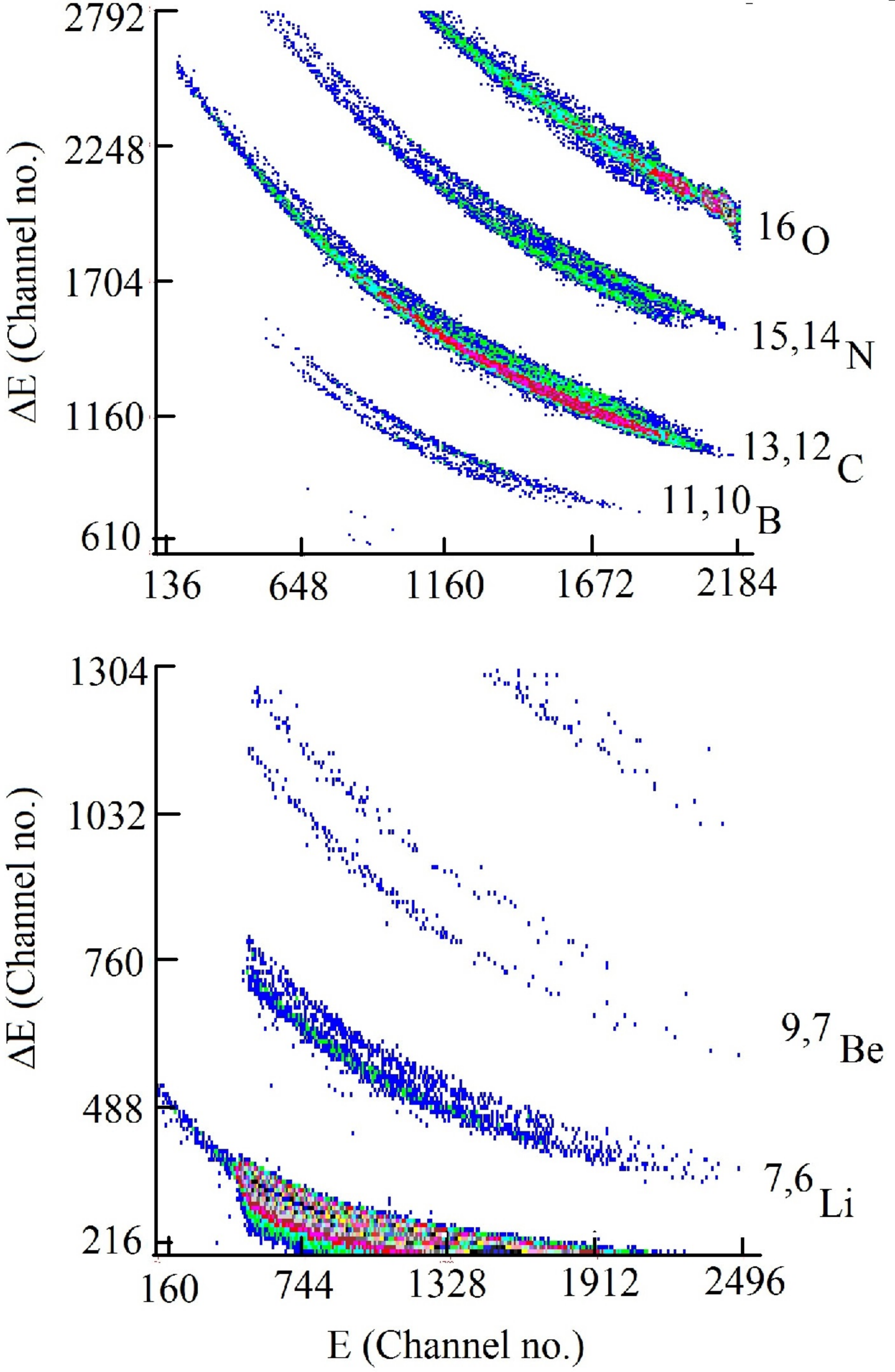}
\caption{(Color online)
A typical two-dimensional $\Delta E$-$E$ spectrum from $^{27}$Al($^{16}$O,\,$x$)
reactions at $E_{\rm lab}=134$~MeV showing the various projectile-like fragments:
(upper panel) oxygen, nitrogen, carbon, and boron isotopes; (lower panel) beryllium
and lithium isotopes.}
\label{2D}
\end{center}
\end{figure}

The experiment was performed with $^{16}$O-ion beam at an incident energy of
$E_{\rm lab}=134$~MeV obtained from the Pelletron-LINAC accelerator facility, Mumbai.
The energy uncertainty of LINAC beam was $\pm0.5$~MeV. The target used was
$^{27}$Al foil of thickness 540~$\mu$g/cm$^2$. Reaction products were detected
using six silicon surface barrier (SSB) detector telescopes in $\Delta E$-$E$ configuration
mounted on two movable arms inside the 1.5-diameter General Purpose Scattering Chamber
in the LINAC beam hall. A typical thickness of $\Delta E$ detectors was $\approx40$~$\mu$m,
while $E$ detectors were of about 1-mm thick. The relative solid angle between the
telescopes was measured by taking data at overlapping angles. The angular resolution
of the telescopes was less than 1.4$^{\circ}$. Data were stored as two-dimensional
$\Delta E$-$E$ spectra and a typical spectrum is presented in Fig.~\ref{2D}, showing
a good charge and mass separation of the projectile-like fragments (PLFs). The reaction
products were identified following the standard particle identification (PI) technique
as described in our earlier study \cite{bjroyPRC2015}.

\begin{figure}[t]
\begin{center}
\includegraphics[width=80mm]{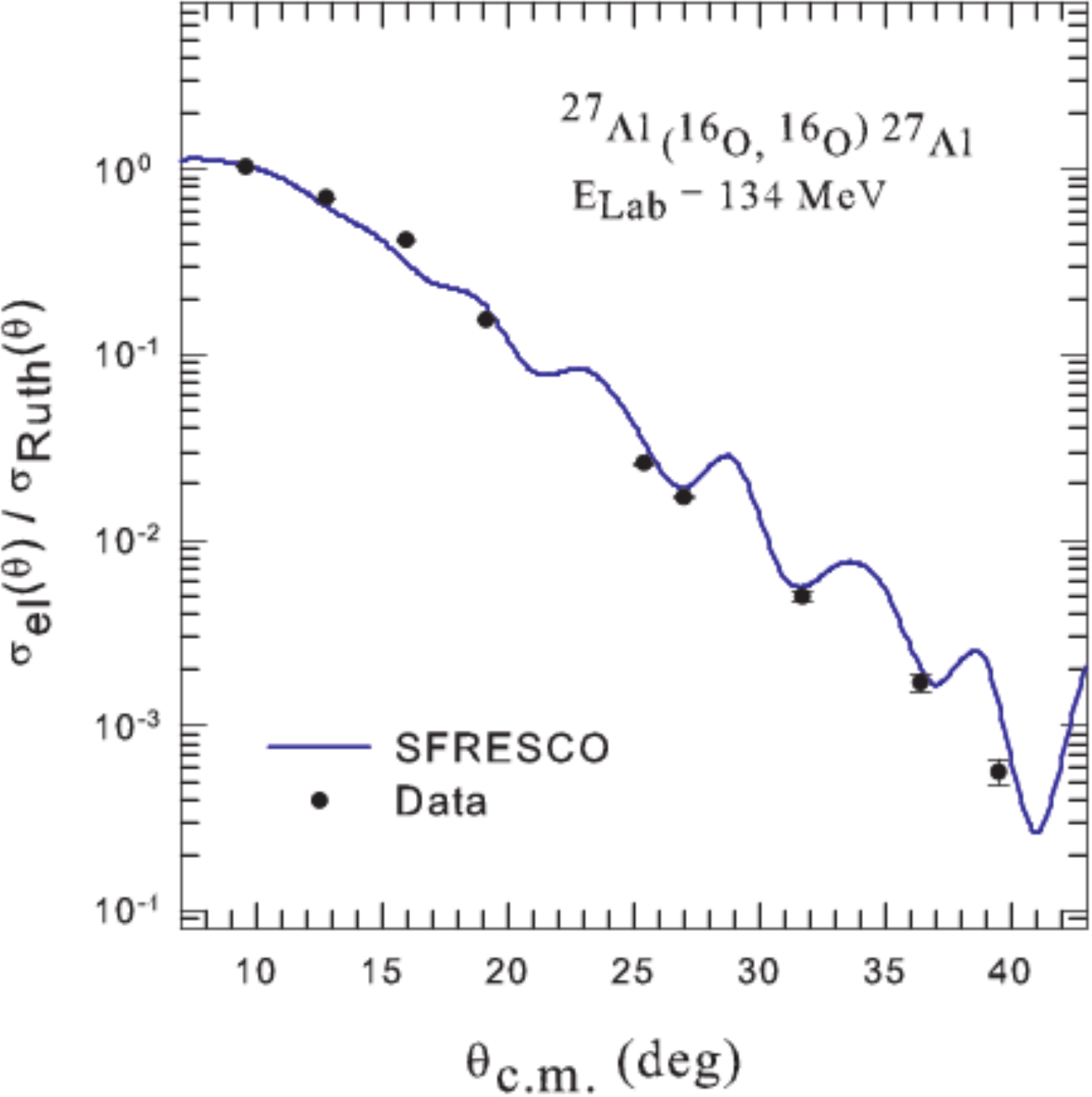}
\caption{(Color online) The ratio of the elastic scattering to the Rutherford cross
sections for the $^{16}$O+$^{27}$Al reaction at $E_{\rm lab}=134$~MeV plotted
as a function of the scattering angle in the center-of-mass frame. Fitted result
by the optical model calculation with the SFRESCO code \cite{Thom01} is shown
by a blue solid curve. The potential parameters obtained from the fitting to the
data are given in Table~\ref{table1}.}
\label{elast}
\end{center}
\end{figure}

\begin{table}[b]
\begin{center}
\caption{Potential parameters for the $^{16}$O+$^{27}$Al system
obtained from the optical-model analysis of the measured elastic-scattering
cross section using the computer program SFRESCO \cite{Thom01}. The
cumulative reaction cross section, $\sigma_{\rm R}$, is also shown. \label{table1}}
\begin{tabular}{c|c|c}
\hline\hline
Potential parameters & $E_{\rm lab}=42$~MeV \cite{Eisen(1977)} & $E_{\rm lab}=134$~MeV \\
\hline
$V_o$ (MeV) & 60.0 & 67.46    \\
$r_o$ (fm) & 1.265 & 0.9   \\
$a_o$ (fm) & 0.472 & 0.55    \\
$W$ (MeV)    & 9.0 & 10.0  \\
$r_i$ (fm) & 1.265 & 1.08    \\
$a_i$ (fm) & 0.472 & 0.71  \\
\hline
$\sigma_{\rm R}$   & 978.16~mb  & 1379.3~mb \\
\hline\hline
\end{tabular}
\end{center}
\end{table}

\begin{figure}[b]
\begin{center}
\includegraphics[width=80mm]{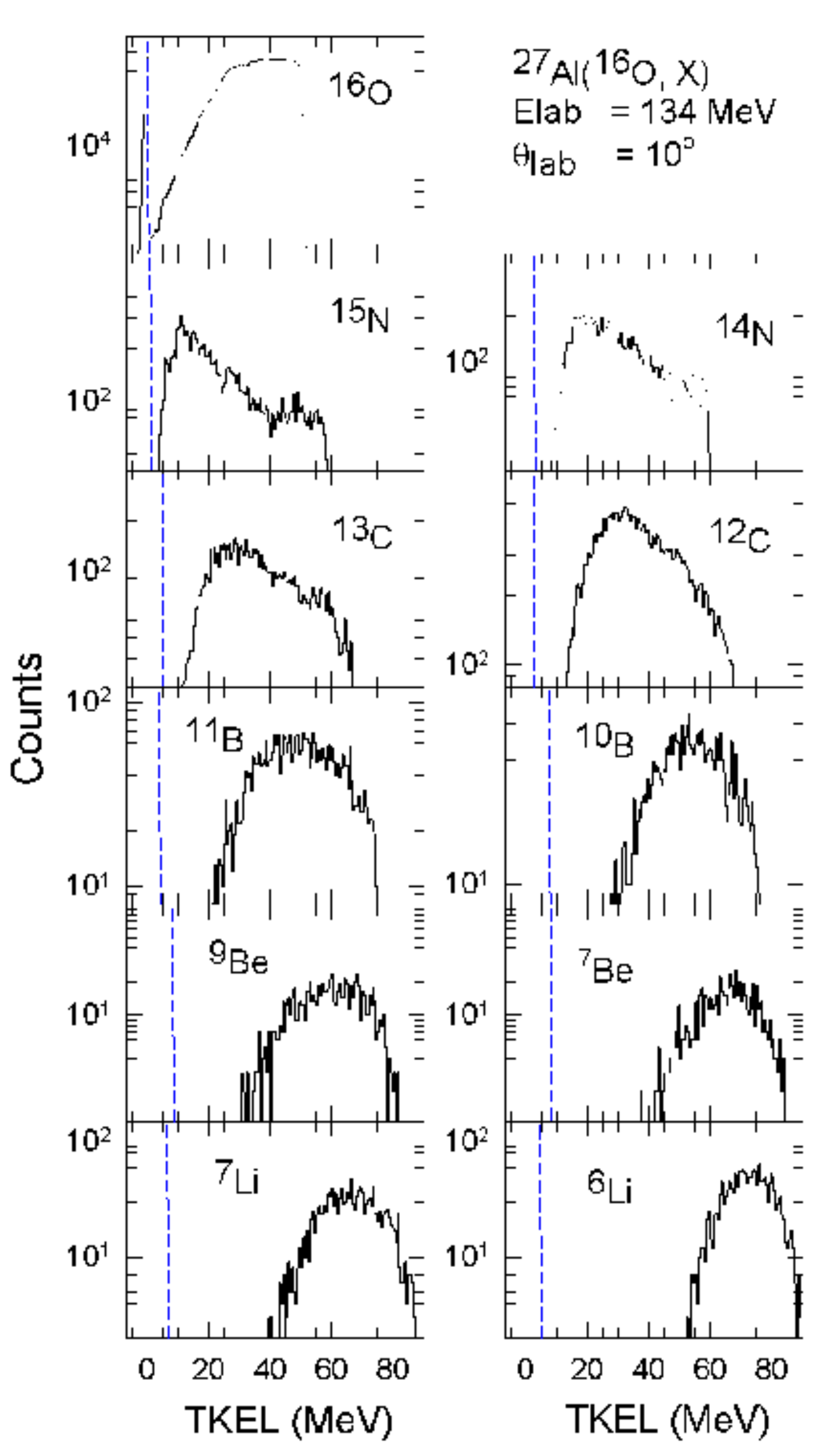}
\end{center}\vspace{-3mm}
\caption{(Color online)
Experimental total kinetic energy loss (TKEL) distributions (histogram plot)
for various projectile-like fragments in the $^{16}$O+$^{27}$Al reaction
at $E_{\rm lab}=134$~MeV, for a fixed scattering angle of $\theta_{\rm lab}
=10^\circ$ ($\theta_{\rm c.m.}=15.9^\circ$). The blue vertical dashed lines
indicate the position of ground-to-ground state $Q$ values.}
\label{tkel}
\end{figure}

\begin{figure*}[t]
\begin{center}
\includegraphics[width=150mm]{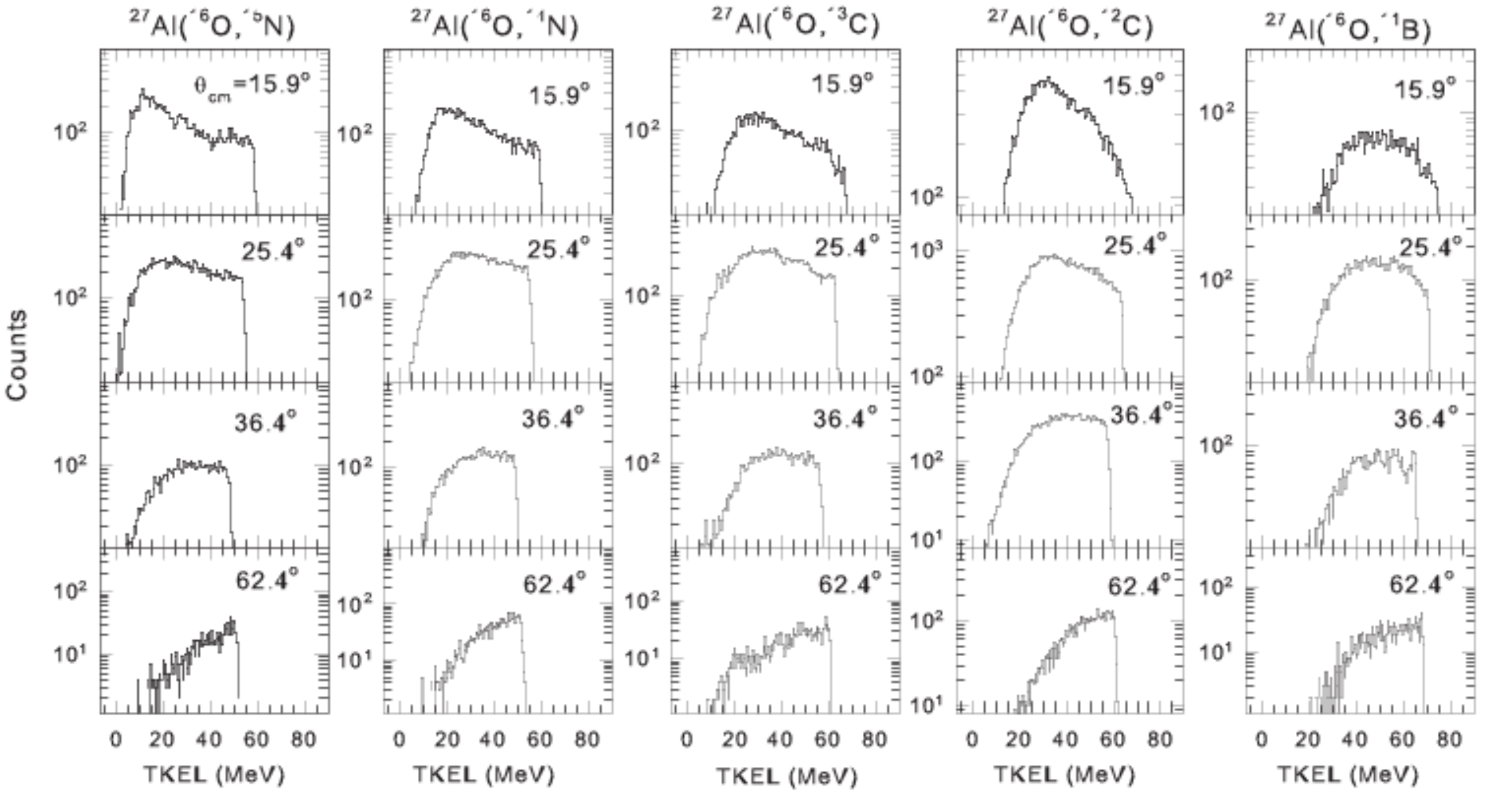}\vspace{2mm}
\includegraphics[width=150mm]{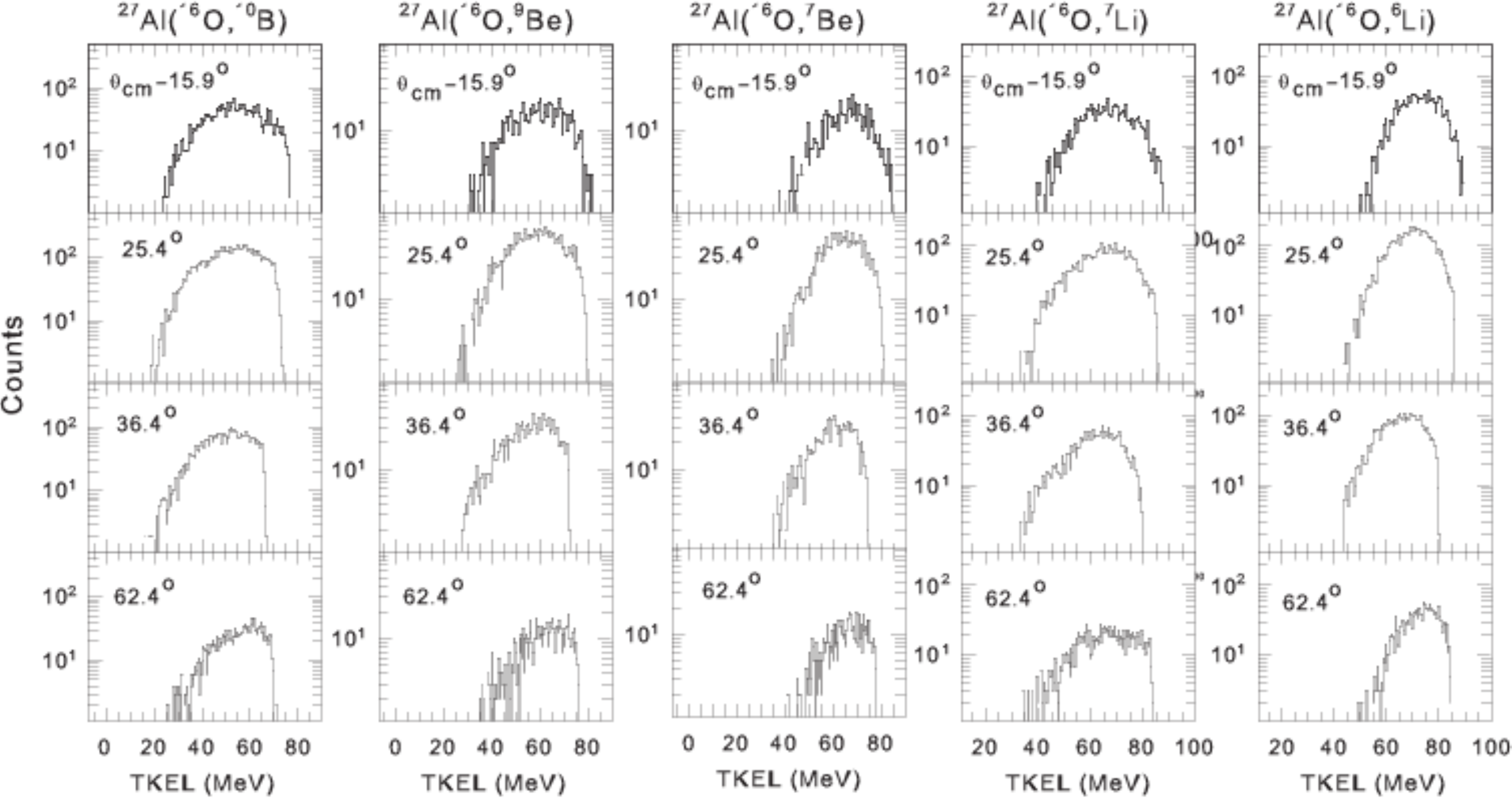}
\end{center}\vspace{-4mm}
\caption{The experimental total kinetic energy loss spectra of
$^{15}$N, $^{14}$N, $^{13}$C, $^{12}$C, and $^{11}$B (top row, from left to right) and
$^{10}$B, $^{9}$Be, $^{7}$Be, $^{7}$Li, and $^{6}$Li (bottom row, from left to right)
reaction products from the $^{16}$O+$^{27}$Al reaction at $E_{\rm lab}=134$~MeV.
Note that the vertical axis was adjusted case by case for better visibility. Note also that
the data shown in the top row ($\theta_{\rm c.m.}=15.9^\circ$) are the same data
as shown in Fig.~\ref{tkel}.}
\label{tkelVStheta}
\end{figure*}

The elastic-scattering angular distribution has been measured and are shown in Fig.~\ref{elast}.
The data are plotted along with statistical errors and in most of the cases the error bars are
within the data symbol. The absolute cross section was obtained by measuring the target
thickness and detector solid angle. Target thickness was measured by alpha-energy loss
method that introduces an error of about 2\% in the absolute cross section. In addition
the error due to the solid angle measurement was $\sim$1\%. The angular distribution has
been analyzed using the optical-model search program SFRESCO \cite{Thom01}.
A volume Woods-Saxon form is used for the real and imaginary parts of the potential.
The optical-model potential parameters of the same system studied at an incident energy of
42~MeV \cite{Eisen(1977)} were used as starting parameters. The obtained best-fit potential
parameters are listed in Table~\ref{table1}. The cumulative reaction cross section, $\sigma_{\rm R}$,
is also shown in the table.

The $Q$-value spectra and angular distributions have been measured for reaction channels
 $^{27}$Al($^{16}$O,\;$x$), where $x$ = $^{15,14}$N, $^{13,12}$C, $^{11,10}$B, $^{9,7}$Be,
and $^{7,6}$Li, which correspond to a mass-number difference of up to $\Delta A=A_{\rm PLF}-A_{\rm Projectile}=-10$.
It is to mention that the pure neutron transfer channels, though visible at some of the angles,
were not clearly separated from the $^{16}$O band in most of the cases and hence we could
not study the TKEL spectra and angular distributions for those reactions. Also, for the spectra
below lithium ($Z=3$), mass separation was not possible, and the energy spectrum for $Z=2$
reaction channel had significant fold back and overlaps with $Z=1$ channel. It was therefore
not possible to extract cross section for $^4$He (alpha) production.

The experimental TKEL spectra at $\theta_{\rm lab}=10^{\circ}$ ($\theta_{\rm c.m.}=15.9^\circ$)
for various channels are shown in Fig.~\ref{tkel}. The TKEL are derived assuming a pure binary process
\cite{montanari01,kim01,Rehm(1991),Corradi(2002)}. For the transfer reactions, as the number of
transferred nucleons increases, there is a gradual shift of the centroid of energy-loss spectra toward the
larger TKEL and reaches to as large as $\sim$ 70~MeV for the $^{27}$Al($^{16}$O,\,$^{6}$Li) channel.
We note that even for the inelastic channel without nucleon transfer ($^{16}$O,\,$^{16}$O$^*$),
a substantial energy loss up to about 50~MeV has been observed (see, the left-top panel of Fig.~\ref{tkel}).
The sharp falloff above 50~MeV corresponds to low energy part of $^{16}$O that gets stopped in the
$\Delta E$ detector. Population of the bulk of events with large energy losses (corresponds to higher
excitation energy) may be associated with high density of excited states, which was observed in several
other studies, \textit{e.g.}, in the $^{27}$Al($^{16}$O,\,$^{16}$O$^*$) inelastic scattering at 280~MeV
\cite{Cappuzzello(2016)} and in the $^{27}$Al($\alpha$, $\alpha'$) scattering \cite{Mayer01,Kiss01}.
The strong excitation in this region, as mentioned in those references, could be due to contributions from
the isoscalar giant quadrupole resonance mode \cite{Kiss01} and octapole excitations \cite{Mayer01}
of $^{27}$Al. There could also be contributions from $^{16}$O particles originated from decay of excited
ejectiles created in pick-up reactions like $^{17}$O, decaying into $^{16}$O\,+\,$n$ \cite{Cappuzzello(2016)}.

To get better understanding of the reaction mechanism, scattering angle dependence of the TKEL spectra
has also been examined. In Fig.~\ref{tkelVStheta}, we show the TKEL spectra of various reaction
products for several selected scattering angles, $\theta_{\rm c.m.}=15.9^\circ$, 25.4$^\circ$, 36.4$^\circ$,
and 62.4$^\circ$. By comparing the TKEL spectra of various transfer channels for each fixed scattering
angle (on the same row), one can again see the gradual shift toward larger TKEL, as the number of transferred
nucleons increases. For channels with relatively small number of transferred nucleons (\textit{i.e.} $^{14,15}$N,
$^{12,13}$C), we find a clear angular dependence of the TKEL spectra. Namely, for those channels low-TKEL
components dominate at forward angles, but the low-TKEL components decreases rapidly as the scattering angle
increases. On the other hand, for processes with a reaction product of $^{10,11}$B, $^{7,9}$Be, and $^{6,7}$Li,
we observe somewhat smaller angular dependence, always dominated by large-TKEL components.
It may indicate that those reaction products are associated with deep-inelastic reactions and are
affected substantially by secondary de-excitation processes.

\begin{figure}[t]
\begin{center}
\includegraphics[width=\columnwidth]{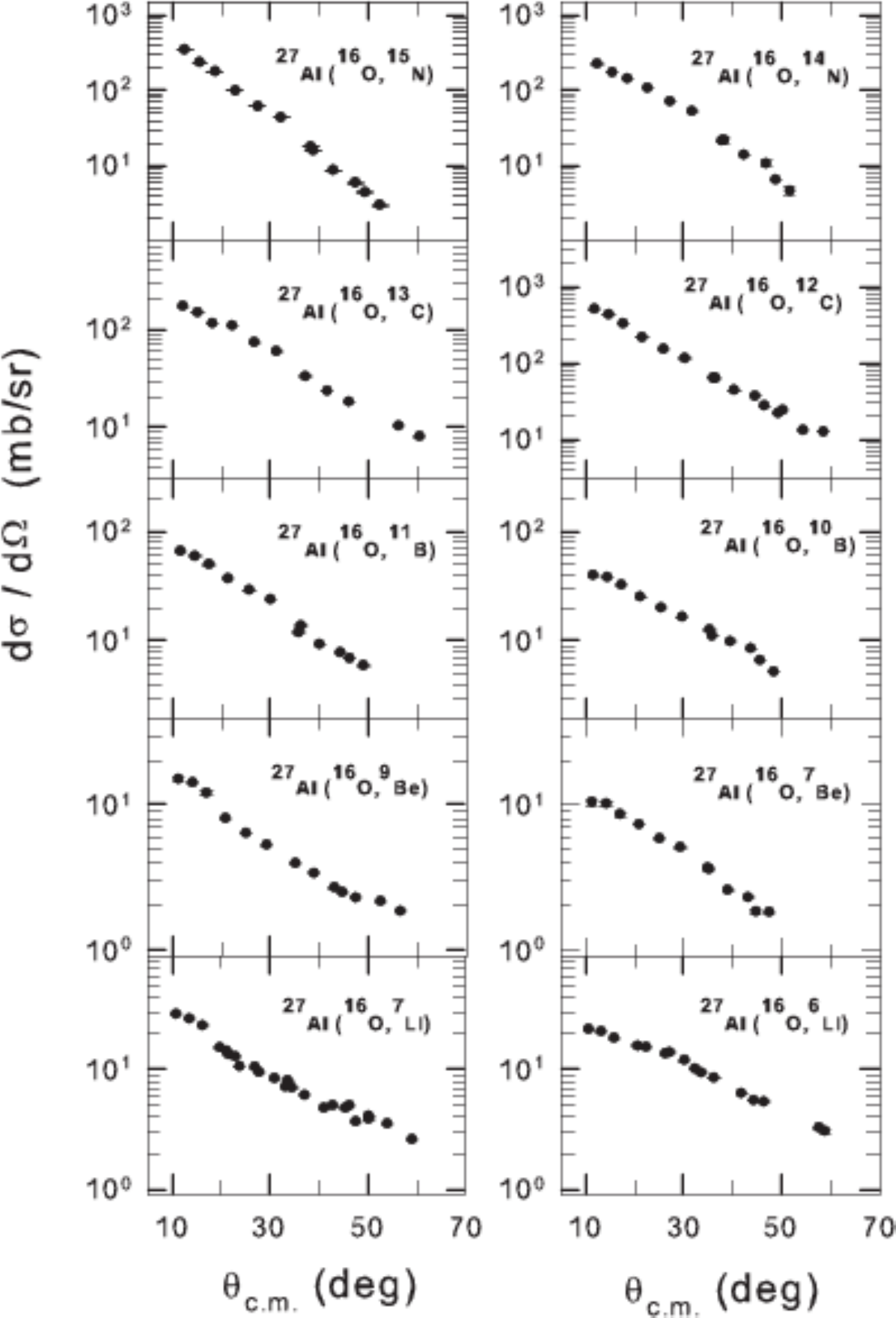}
\caption{
The measured $Q$-value-integrated angular distributions for various channels
in the $^{16}$O+$^{27}$Al reaction at $E_{\rm lab}=134$~MeV. The errors
indicate the statistical error.
}
\label{o-n-ang}
\end{center}
\end{figure}

The $Q$-value-integrated angular distributions for various reaction channels
have been obtained by appropriate selection of mass and charge of the reaction
products and are shown in Fig.~\ref{o-n-ang}. The angular distributions are forward
peaked, consistent with earlier studies for the same system at energies well above the
Coulomb barrier \cite{Cormier(1976),Mikumo(1980)}. The strong forward-peaked nature
of the multinucleon transfer angular distributions, in contrast to the bell-shaped angular
distributions which are typical characteristic of quasi-elastic processes, is an indication
of the increasing importance of nuclear effects and dissipative processes that
occur at shorter inter-nuclear distance \cite{bjroy(2002)}.

\begin{figure*}[t]
   \begin{minipage}{\columnwidth}
        \begin{center}{\hspace{8mm}\bf (a) Primary products}
        \includegraphics[width=83mm]{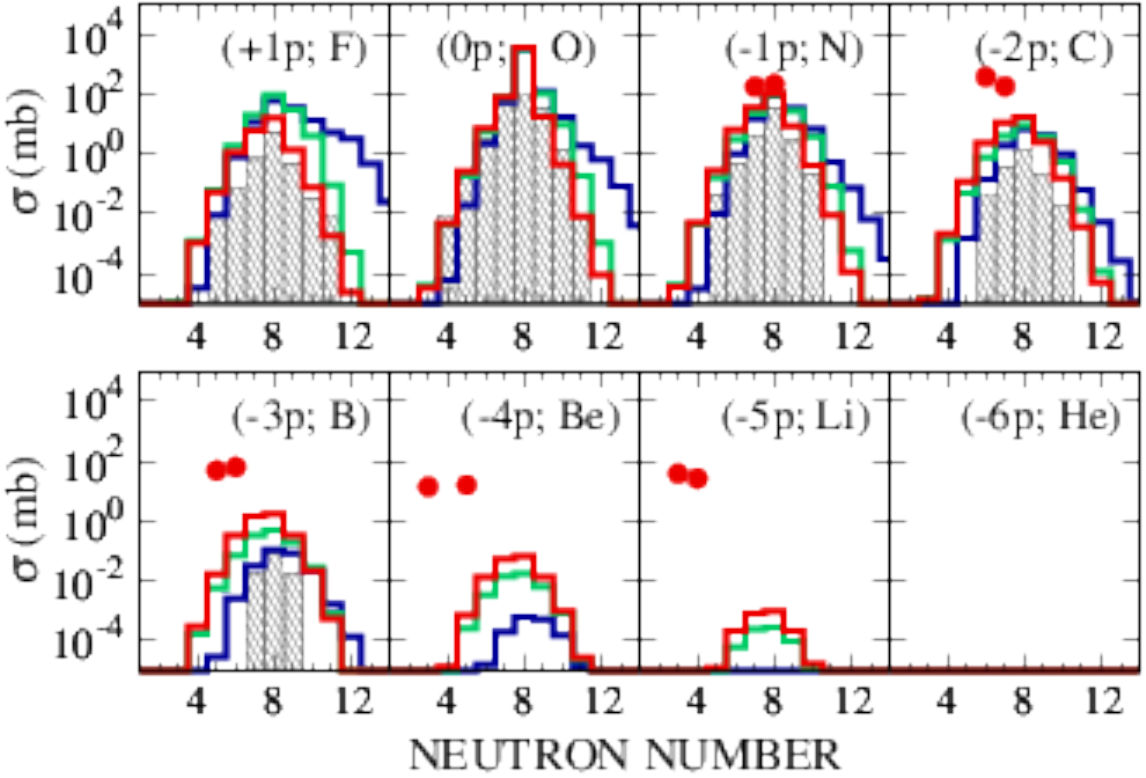}
        \end{center}
   \end{minipage}
   \hspace{1mm}
   \begin{minipage}{\columnwidth}
        \begin{center}{\hspace{8mm}\bf (b) Secondary products}
        \includegraphics[width=83mm]{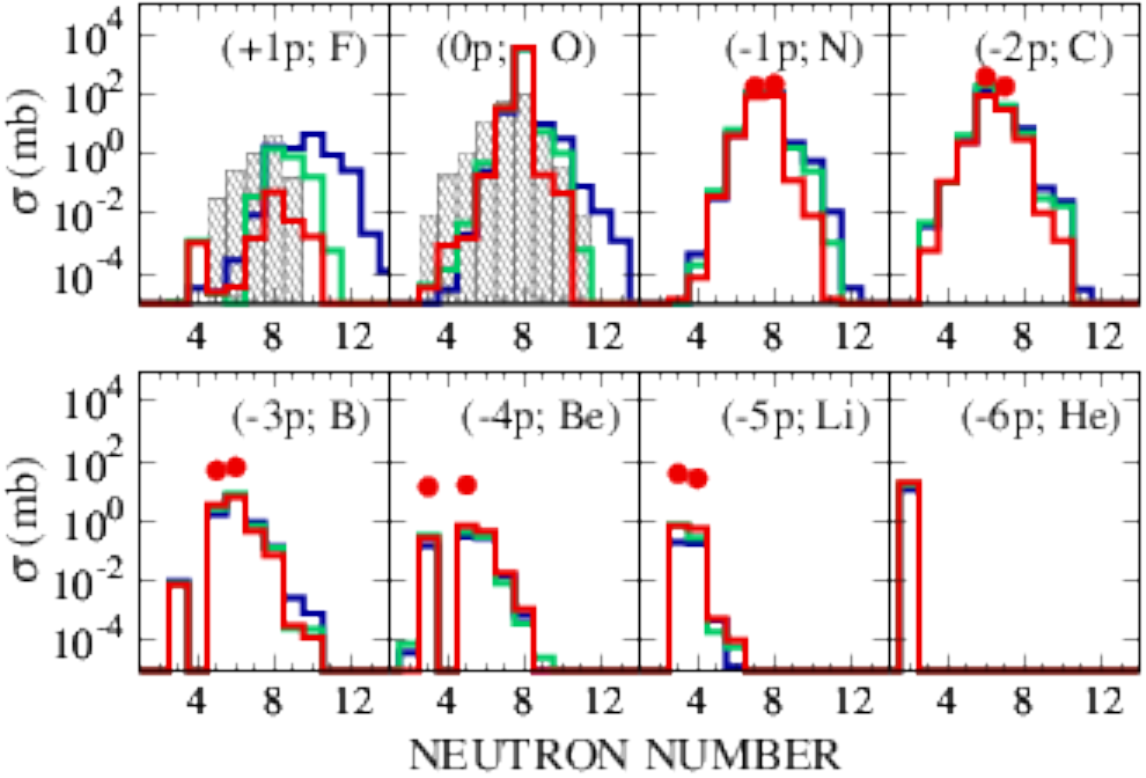}
        \end{center}
   \end{minipage}
   \caption{(Color online)
   The $Q$-value- and angle-integrated isotope production cross sections for various
   proton-transfer channels in the $^{16}$O+$^{27}$Al reaction at $E_{\rm lab}=134$~MeV.
   The change in the number of protons compared with the projectile ($Z=8$) is indicated
   as ($\pm x$p;~X), where X stands for the corresponding element. Red filled circles show the
   experimental data. The results of TDHF(+GEMINI) calculations are shown by histograms.
   Calculated results for primary (secondary) products are shown in the left (right) figure.
   {\footnotesize GRAZING} results \cite{GRAZING-code} are also shown by gray
   shaded histograms for comparison.
   }
   \label{FIG:sigmatot}
\end{figure*}

The $Q$-value- and angle-integrated isotope production cross sections for various
transfer processes have been obtained by fitting the angular distribution by an exponential
function $d\sigma/d\theta \propto \exp[-\alpha\theta]$ (as suggested in Refs.~\cite{Galin(1976),Lefort(1978)})
and integrate over whole angular range, 0 to 180$^\circ$. The fitting procedure introduced an
uncertainty of 9--13\% in the final value of the cross section in most of the reaction
channels except for $^9$Be and $^{13}$C for which a slightly larger error
($\approx$ 16--17 \%) was obtained. We note that we also tried a fitting with
$d\sigma/d\theta \propto \exp[-\alpha\theta]/\sin\theta$ (used in Ref.~\cite{Mikumo(1980)}),
but the latter was giving somewhat poor fit in some of the reaction channels and resulted
in large errors (more than 30\% error in some cases). Therefore, we have decided to
use the fit with the simple exponential function. The integrated cross sections are presented
in Sec.~\ref{sec:comparison} (Fig.~\ref{FIG:sigmatot}), where we compare the data with
theoretical calculations.

\section{TDHF ANALYSIS}\label{sec:TDHF}

To gain deeper insight into the reaction mechanism, we have performed TDHF
calculations for the $^{16}$O+$^{27}$Al reaction. We used a parallel computational
code developed by Sekizawa and Yabana \cite{KS_KY_MNT}. The code has been tested
and successfully applied for various systems \cite{KS_KY_MNT,KS_KY_PNP,MyPhD,bjroyPRC2015,
KS_KY_Ni-U,KS_SH_Kazimierz,Sekizawa(2017)}. The details of the calculations can be found
in the references given above. Here we provide brief information relevant to the present
analysis. Details of the TDHF theory and its application to nuclear dynamics can be found in
Refs.~\cite{Negele(review),Simenel(review),Nakatsukasa(PTEP),Sky3D,Nakatsukasa(review)}.

In the TDHF approach, internal degrees of freedom are described microscopically from
nucleonic degrees of freedom; whereas the collective motions of two colliding nuclei are
described semiclassically through dynamics of a time-dependent mean-field potential
generated by the interaction among all the nucleons. The theory can describe important
features of the low-energy heavy ion reaction, such as the Pauli principle among nucleons,
shell effects, energy dissipations, nucleon exchanges, and shape evolution dynamics in the
composite system formed in the course of the collision, on the same footing. The only input
to the calculations is an effective nucleon-nucleon interaction (or an energy density functional,
EDF), which is determined so as to reproduce known properties of finite nuclei and nuclear matter.
Thus, in the TDHF approach, there is no adjustable (empirical) parameter specific to the reaction.

\begin{figure}[t]
   \begin{center}
   \includegraphics[width=4.5cm]{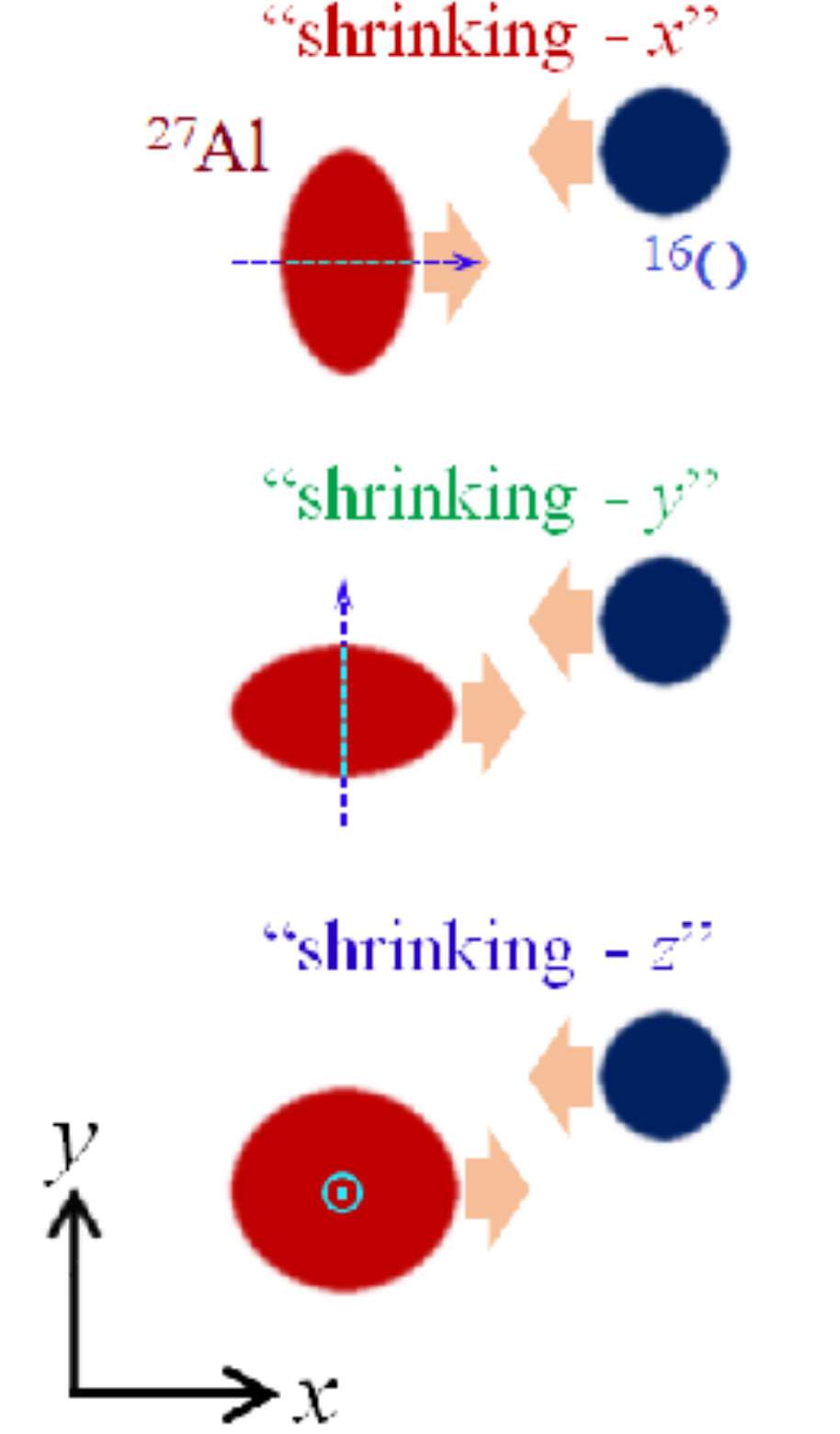}
   \end{center}\vspace{-3mm}
   \caption{(Color online)
   A schematic illustration of the initial orientations of $^{27}$Al in the TDHF calculations.
   Red (blue) disc represents cross sections of the density of $^{27}$Al ($^{16}$O) nucleus
   in the reaction plane ($xy$ plane). The incident direction is indicated by thick arrows which
   are parallel to the $x$ axis (at infinitely large distance), and the impact parameter vector
   is parallel to the $y$ axis. By a blue dashed arrow or a circle attached to $^{27}$Al, the
   direction of shrinking is represented. This direction would correspond to the symmetry axis
   if $^{27}$Al were purely deformed in an oblate shape.
   }
   \label{FIG:orientations}
\end{figure}

For the EDF, the Skyrme SLy5 parameter set \cite{Chabanat} was used. Static
Hartree-Fock (HF) calculations were performed in a cubic box with 24~fm on each side.
The mesh spacing was set to 0.8~fm for both static and dynamic calculations. The HF
ground state of $^{16}$O is of spherical shape, whereas that of $^{27}$Al turned out
to be of a triaxial shape with $\beta\simeq0.27$ and $\gamma\simeq48^\circ$. For
the $^{27}$Al nucleus, a large oblate deformation is expected \cite{FRDM(2012)}.
Since the deformation is quite large, TDHF calculations were performed taking three
initial orientations of $^{27}$Al. As the shape looks more or less like an oblate shape,
those three orientations will be labeled according to the direction of ``shrinking" of
the total density, which would correspond to the symmetry axis if the nucleus were
purely deformed in an oblate shape (see Fig.~\ref{FIG:orientations}). In the TDHF
calculations, we set the collision axis to the $x$-direction and the impact parameter
vector parallel to the $y$ axis. Thus the reaction plane corresponds to the $xy$ plane.
A three-dimensional box of $60\,{\rm fm}\times52\,{\rm fm}\times24\,{\rm fm}$
was used without symmetry restrictions. The initial separation distance between the
projectile and target nuclei was set to 24~fm. TDHF calculations were performed for
an impact parameter range of $[0,12]$ (fm). The minimum impact parameter for
binary reactions, $b_{\rm min}$, inside which fusion reactions take place was searched
by repeating TDHF calculations with an 0.001-fm impact parameter step. The very small
impact parameter step was chosen so as to capture a rapid change of the reaction
dynamics in the vicinity of the fusion threshold. From the results, we find $b_{\rm min}
=7.025$~fm, 6.348~fm, and 6.899~fm for the shrinking-$x$, $y$, and $z$ cases,
respectively. Inside this impact parameter ($b=0,1,\dots,6$, and $b_{\rm min}-0.001$~fm),
fusion reactions were observed, where a mononuclear composite system persists in
a compact shape within the simulation time of more than 26~zs (1~zs~$=10^{-21}$~sec).
In this context we note that the fusion excitation function for the present system
in the energy range $E_{\rm lab}=25$--200~MeV was studied in the past \cite{Devi(1981)}
within TDHF, where no fusion was observed at small impact parameters (the so-called
fusion window) for  $E_{\rm lab}>100$~MeV that might be due to the various approximations
that were used in those calculations. Coming back to the present TDHF calculations,
the time evolution for the binary reactions ($b\ge b_{\rm min}$) was continued until
the relative distance between the reaction products reaches 26~fm. Various reaction
outcomes were then computed from TDHF wavefunctions after the collision with
well-separated reaction products.

\begin{figure}[t]
   \begin{center}
   \includegraphics[width=6.5cm]{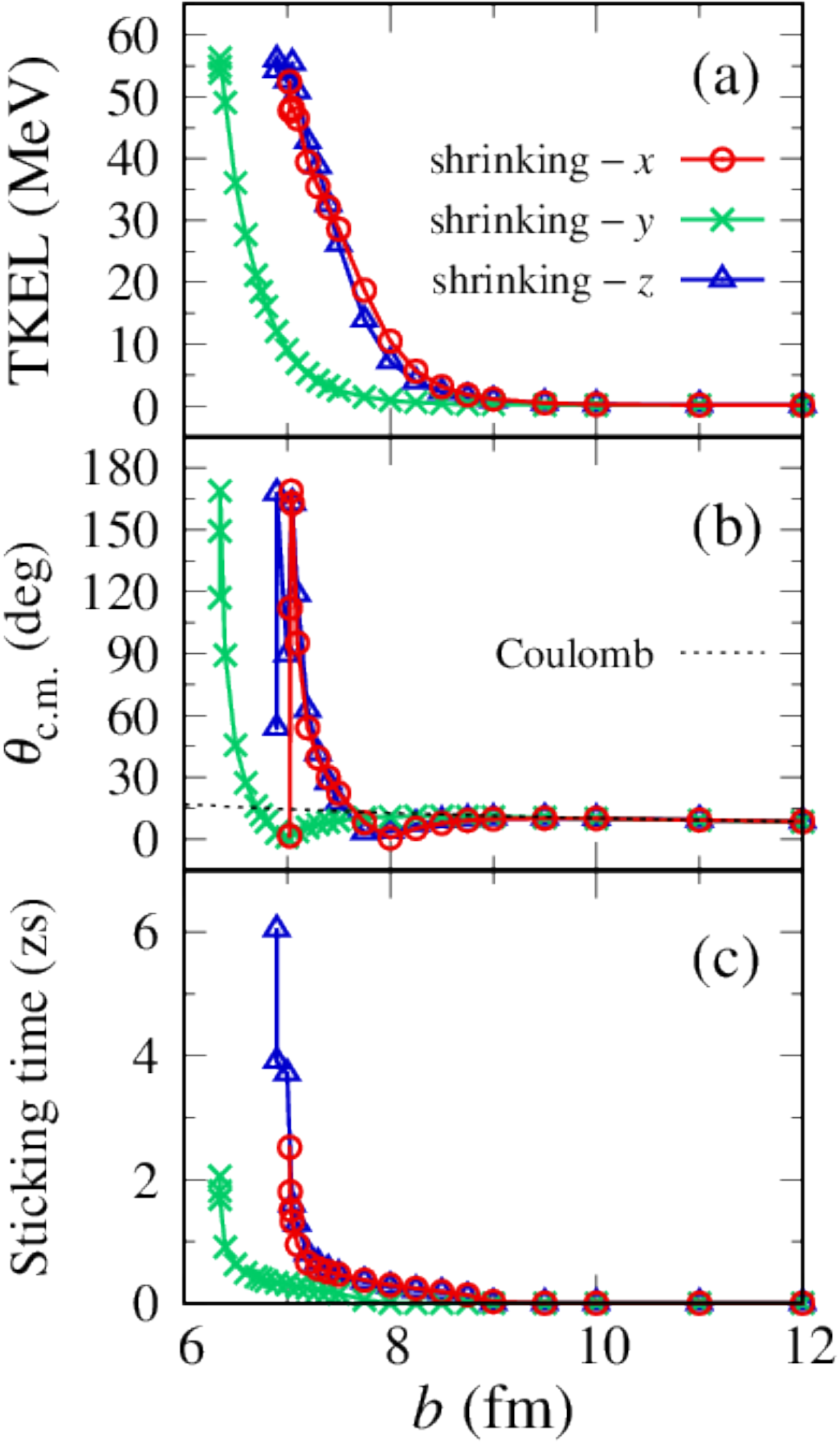}
   \end{center}\vspace{-4mm}
   \caption{(Color online)
   Results of the TDHF calculations for the $^{16}$O+$^{27}$Al reaction
   at $E_{\rm lab}=134$~MeV. (a) Total kinetic energy loss (TKEL), (b)
   scattering angle in the center-of-mass frame, and (c) sticking time are
   shown as a function of the impact parameter, $b$. In (b), the scattering
   angle for the Coulomb trajectory is indicated by a dotted curve. In (c),
   the sticking time is shown in zeptoseconds ($1$~zs~$=10^{-21}$~sec).
   }
   \label{FIG:TKEL-theta}
\end{figure}

First, let us examine global features of the reaction dynamics. From the
dynamics of the mean-field potential, average (most probable) reaction
trajectories can be evaluated. In Fig.~\ref{FIG:TKEL-theta}, we show TKEL
in (a), the scattering angle in the center-of-mass frame in (b), and the sticking
time in (c) as a function of the impact parameter. Red open circles, green crosses,
and blue open triangles show the results associated with the different initial orientation
of $^{27}$Al, corresponding to the shrinking-$x$, $y$, and $z$ cases, respectively
(\textit{cf.} Fig.~\ref{FIG:orientations}). Henceforth, the same colors and symbols
will be used to indicate these orientations in the figures. In (b), the scattering angle
for the Coulomb trajectory is indicated by a black dotted curve. The sticking time
is the time duration in which the lowest density between two colliding nuclei exceeds
a critical value, $\rho_c=0.01$~fm$^{-3}$.

From the figure, we find that the global features of the reaction dynamics do not
depend much on the initial orientations of $^{27}$Al. When the impact parameter
is relatively large ($b>9$~fm), TKEL is very small and the scattering angle coincides
with the one for the Coulomb trajectory. As the impact parameter decreases ($b<9$~fm),
TKEL increases rapidly, reaching the maximum value of around 55~MeV. The maximum
value of TKEL reasonably agrees with the experimental observation in Figs.~\ref{tkel}
and \ref{tkelVStheta}. In this regime, the density of the colliding nuclei overlaps in the
course of the collision, as can be seen as a gradual increase of the sticking time in (c),
and, as a result, the scattering angle largely deviates from the Coulomb trajectory. We
note that no nuclear rainbow is observed in TDHF, as all trajectories for $b<b_{\rm min}$
resulted in fusion; the large scattering angles merely correspond to large negative deflection angles.
Only a visible orientation dependence seen in Fig.~\ref{FIG:TKEL-theta} is a shift of the
curves toward the smaller impact parameters (about 1~fm) in the shrinking-$y$ case (green crosses),
as compared to the other cases. This difference can be simply understood from the collision
geometry depicted in Fig.~\ref{FIG:orientations}. Since we chose the impact parameter
vector parallel to the $y$ axis, the system requires smaller impact parameters in order
to collide deeply in the shrinking-$y$ case. This explains the orientation dependence
seen in Fig.~\ref{FIG:TKEL-theta}. (A somewhat longer sticking time for $b \approx
b_{\rm min}$ in the shrinking-$z$ case will be discussed later.)

\begin{figure}[t]
   \begin{center}
   \includegraphics[width=6.5cm]{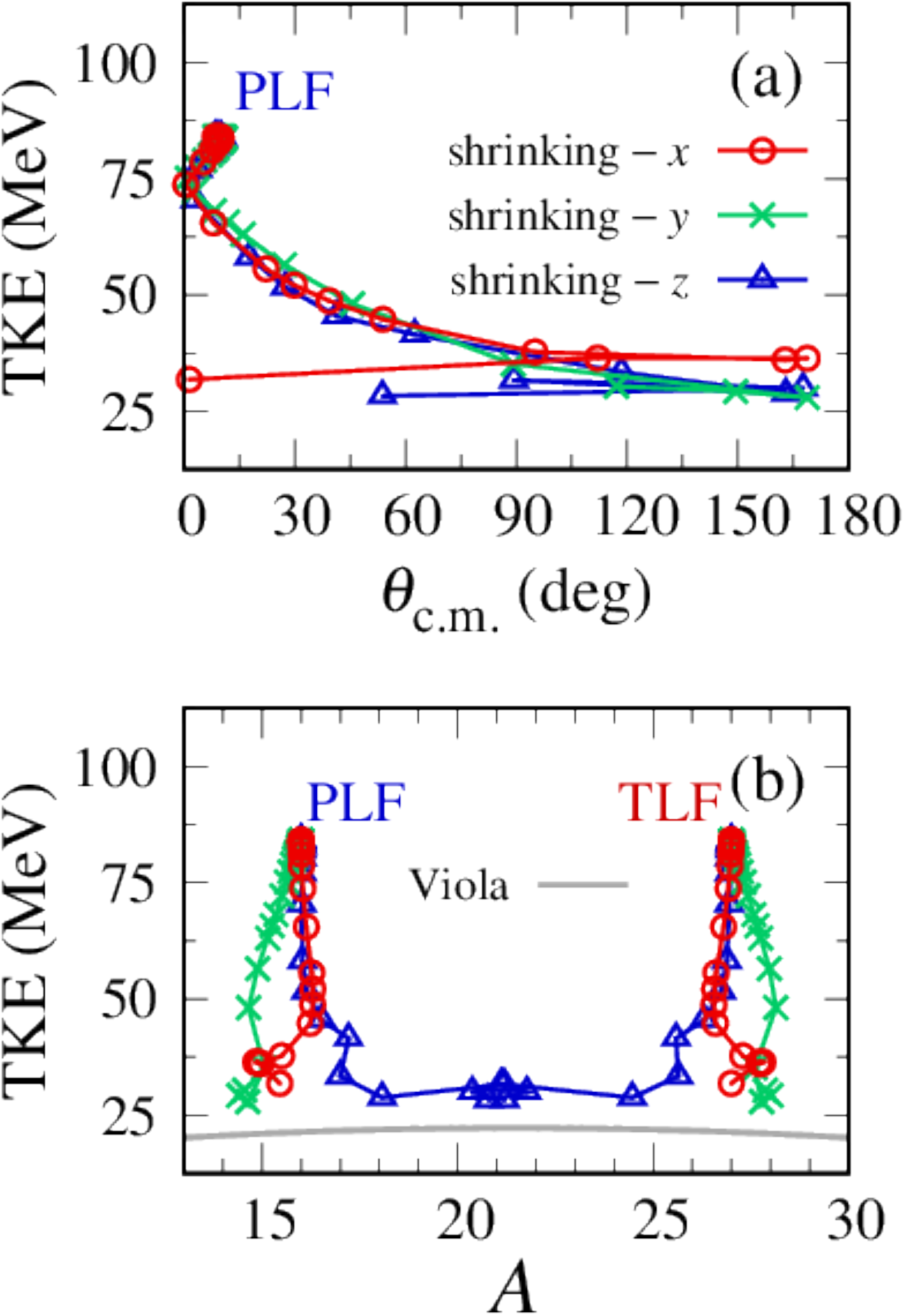}
   \end{center}\vspace{-4mm}
   \caption{(Color online)
   (a) The Wilczy\'nski plot and (b) the TKE-$A$ distribution obtained from
   the TDHF calculations for the $^{16}$O+$^{27}$Al reaction at $E_{\rm lab}
   =134$~MeV.
   }
   \label{FIG:TKE}
\end{figure}

Combining the information of the TKEL and the scattering angle in Fig.~\ref{FIG:TKEL-theta}~(a,\,b),
we make the so-called Wilczy\'nski plot \cite{Wilczynski(1973)} which is shown
in Fig.~\ref{FIG:TKE}~(a). We note that it has been considered that TDHF can
reasonably capture main reaction dynamics in deep-inelastic collisions (\textit{i.e.}
the most probable trajectory in the Wylczy\'nski plot) since the very early stage \cite{Davies(1978),Davies(1979),SandhyaDevi(1981),Davies(1981),Dhar(1984)},
which has been confirmed also in a recent experimental/theoretical study reported
in Ref.~\cite{Williams(2018)}. The plot shows characteristic behavior for deep-inelastic
reactions: a decrease of the scattering angle with larger TKE, followed by a rapid increase
of the scattering angle accompanying large energy losses. At this collision energy substantially
above the Coulomb barrier, a dinuclear system rotates a lot in the reaction plane and it
finally re-separates. The dynamics result in appearance of the fragments in a wide angular
range after the significant energy dissipation, indicating orbiting dynamics of the dinuclear system.

In contrast to the reaction dynamics shown in Figs.~\ref{FIG:TKEL-theta} and
\ref{FIG:TKE}~(a), we find that the fragment masses show noticeable orientation
dependence. It can be clearly seen in Fig.~\ref{FIG:TKE}~(b), where we show
the total kinetic energy \textit{vs.} mass distribution of the reaction products.
For comparison, empirical values from the Viola systematics \cite{Viola0,Viola1,Viola2}
are also shown by a gray solid curve. The two peaks at around $A$ = 16 and 27
correspond to the quasielastic peak for the PLF and the target-like fragment (TLF),
respectively. From the figure, we see that the reaction products tend to be slightly
mass asymmetric in the shrinking-$y$ case (green crosses). On the other hand,
up to certain energy losses (${\rm TKE}\approx40$), the average fragment masses
are nearly constant in the shrinking-$x$ and $z$ cases. As the energy loss increases
further (${\rm TKE}\lesssim40$), however, those two cases deviate each other.
Intriguingly, only in the shrinking-$z$ case (blue open triangles) the composite
system tends to evolve toward the mass symmetry. In this case, the reaction
proceeds on the equatorial side of the oblate-like $^{27}$Al (\textit{cf.}
Fig.~\ref{FIG:orientations}). We note that for all orientations TKE values become
rather close to the Viola systematics, although not fully coincide. The behavior
nicely mimics the one observed in heavy systems.

\begin{figure}[t]
   \begin{center}
   \includegraphics[width=6.5cm]{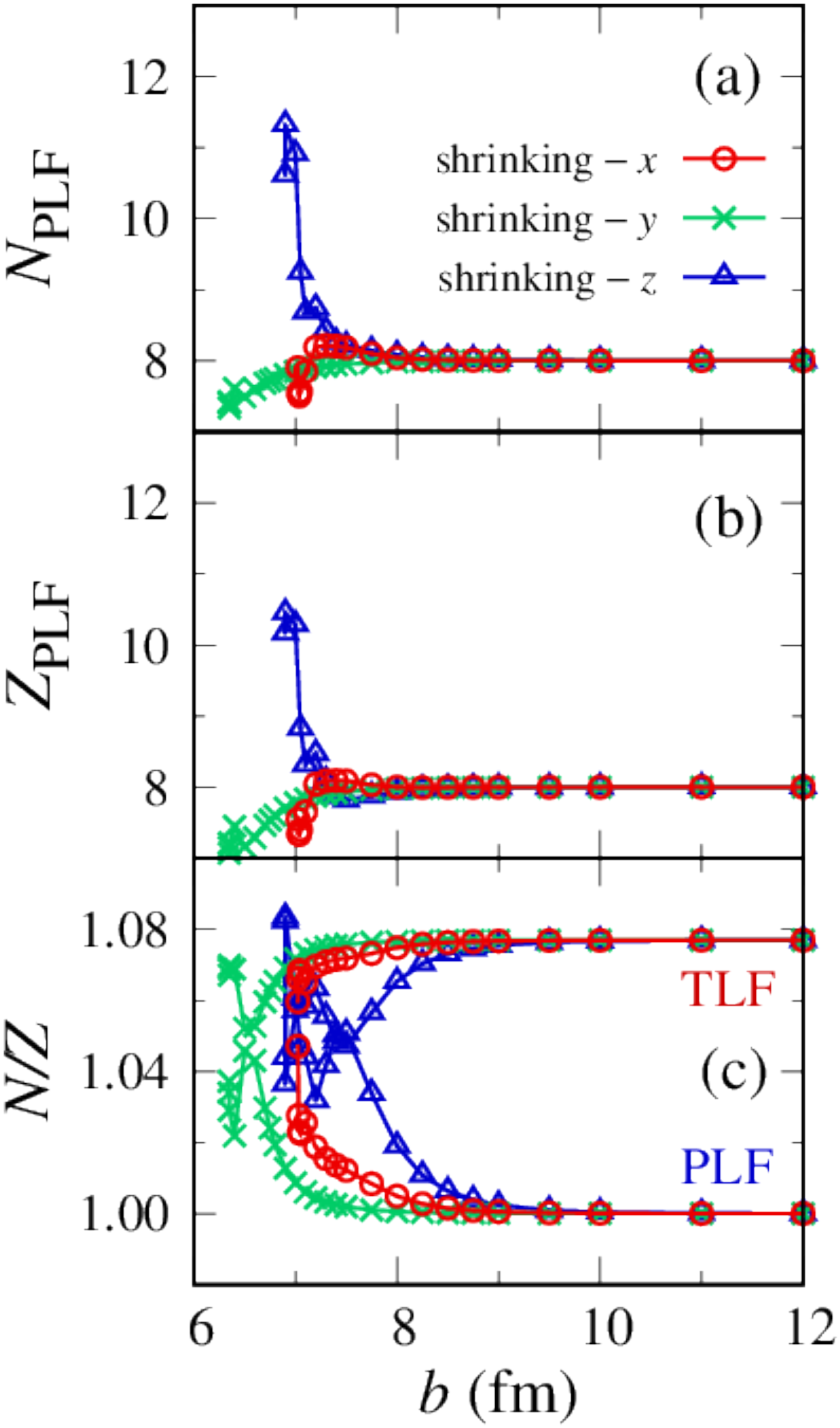}
   \end{center}\vspace{-4mm}
   \caption{(Color online)
   Average numbers of neutrons (a) and protons (b) of the projectile-like fragment (PLF) in $^{16}$O+$^{27}$Al
   reaction at $E_{\rm lab}=134$~MeV are shown as a function of the impact parameter, $b$.
   In (c), the average $N/Z$ ratios of the PLF and the target-like fragment (TLF) are shown.
   }
   \label{FIG:Nave}
\end{figure}

\begin{figure*}[t]
   \begin{center}
   \includegraphics[width=15cm]{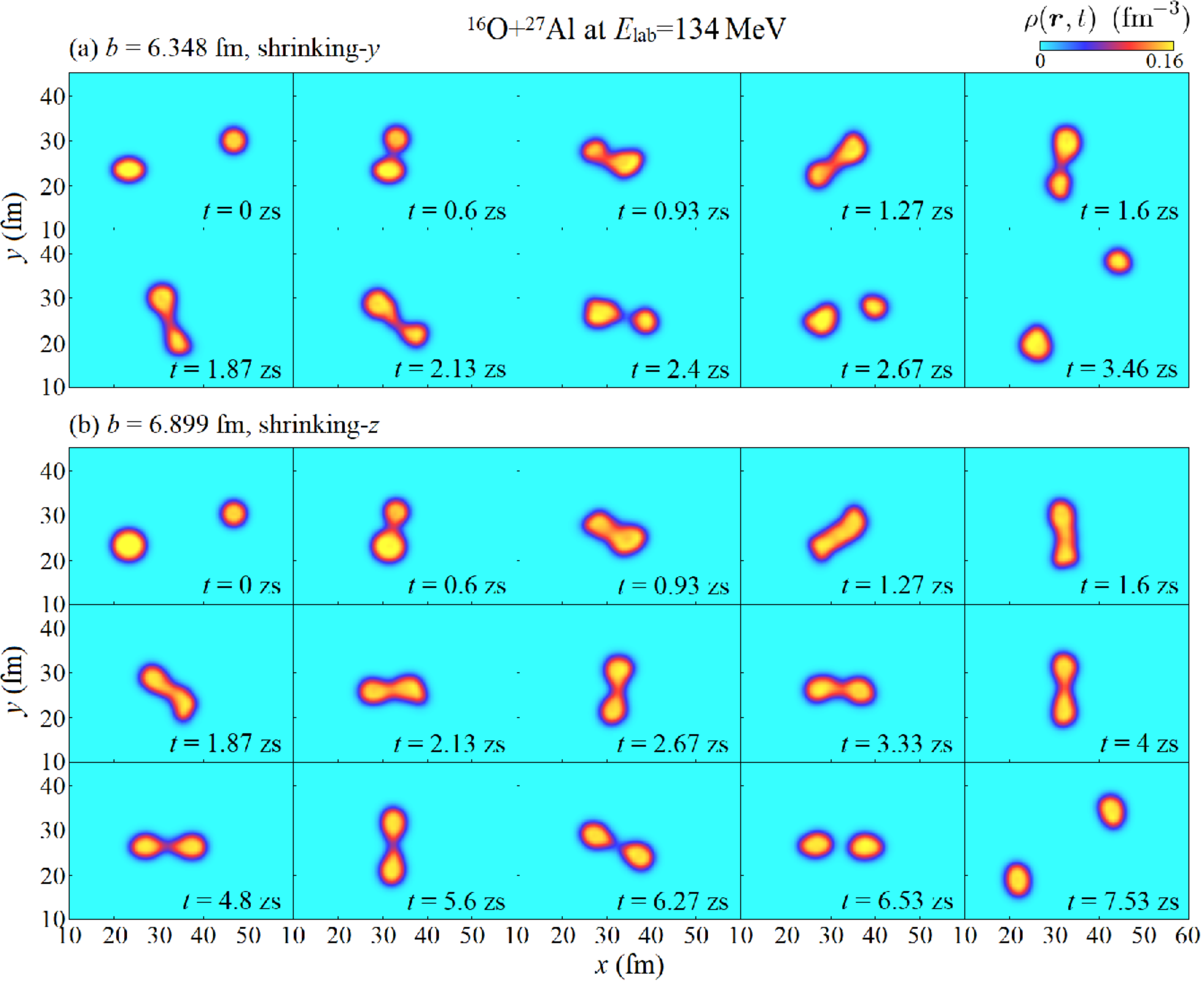}
   \end{center}\vspace{-4mm}
   \caption{(Color online)
   Snapshots of the density of the colliding nuclei in the reaction plane at various times
   obtained from the TDHF calculations for the $^{16}$O+$^{27}$Al reaction at
   $E_{\rm lab}=134$~MeV. In upper panels (a), dynamics for the shrinking-$y$
   case are shown, while those for the shrinking-$z$ case are shown in lower panels (b).
   In each panel, the elapsed time is indicated in zeptosecond ($1$~zs~$=10^{-21}$~sec).
   It is to mention that the whole simulation box ($60\,{\rm fm}\times52\,{\rm fm}$)
   is not shown in the figure.
   }
   \label{FIG:rho(t)}
\end{figure*}

In Fig.~\ref{FIG:Nave}, average numbers of neutrons (a) and protons (b) of the PLF
are shown as a function of the impact parameter. From the figure, we find some orientation
dependence of the transfer processes. In the shrinking-$z$ case, we find that both neutrons
and protons are transferred toward the same direction from $^{27}$Al to $^{16}$O, as
the impact parameter decreases. The maximum values, $N_{\rm PLF}\approx11$ and
$Z_{\rm PLF}\approx10.5$, correspond to the symmetric division of the composite system,
consistent with the observation in Fig.~\ref{FIG:TKE}~(b). On the other hand, in the
shrinking-$y$ case, neutrons and protons are transferred toward the direction slightly
increasing the mass asymmetry. The shrinking-$x$ case exhibits an intermediate behavior,
first few nucleons are transferred from $^{27}$Al to $^{16}$O, then the direction changes,
as the impact parameter decreases. The observed difference may be related to the properties
of single-particle orbitals. In an oblate(-like) nucleus, single-particle orbitals tend to extend
toward the equatorial side, which may prefer nucleon transfer from $^{27}$Al to $^{16}$O.
While along the axis of shrinking, the orbitals tend not to extend, which may prefer transfer in
the opposite direction. A similar trend was observed in our earlier study \cite{bjroyPRC2015},
where a prolately deformed $^{18}$O induced transfer reactions were examined within TDHF.
In Fig.~\ref{FIG:Nave}~(c), average $N/Z$ ratios of the fragments are shown.
Although the $N/Z$ ratios are very close to each other already before the collision,
the reaction tends to proceed toward further charge equilibration. At small impact
parameters close to the fusion threshold, the $N/Z$ ratios fluctuate, because of the
much longer sticking time and complex dynamics in the dinuclear system.

The results of the TDHF calculations indicate that certain mass symmetric fragments
should emerge as a result of deeply inelastic reactions, which exhibit fusion-fission-like
character, \textit{i.e.}, with fully-damped kinetic energy and isotropic angular distribution.
Although it was not possible in the present experimental setup to cover wider angular range
and detect heavier PLFs and TLFs, it is nowadays routinely performed, especially for heavier
systems (see, \textit{e.g.}, Refs.~\cite{Williams(2018),Rietz(2011),Rietz(2013),Wakhle(2014),
Prasad(2017)}, and references therein). Detailed comparisons between experimental and
theoretical data for the mass-angle correlation and the Wilczy\'nski plot would provide us
deeper understanding of the reaction dynamics.

To understand better the observed behavior, we show in Fig.~\ref{FIG:rho(t)}
the time evolution of the density of the colliding nuclei for two representative cases:
the shrinking-$y$ and $z$ cases are shown in panels (a) and (b), respectively.
Both cases correspond to the reaction at the minimum impact parameter for binary
reactions, $b_{\rm min}$. These two cases resulted in production of different nuclei
as mentioned above: the average reaction product was $Z_{\rm PLF}\simeq7.2$
and $N_{\rm PLF}\simeq7.4$ for the shrinking-$y$ case, while it was $Z_{\rm PLF}
\simeq10.2$ and $N_{\rm PLF}\simeq10.6$ for the shrinking-$z$ case. We shall
now examine how the collision dynamics differ in those cases.

In the top row of panels (a) and (b), the initial stage of the reaction is displayed.
As can be seen from the figure, two nuclei start colliding at around $t=0.6$~zs,
then merged deeply due to the strong attractive interaction ($t=0.93$~zs). Because
of the substantial angular momentum brought into the system, an elongated composite
system is created ($t=1.27$~zs). At this stage, one can see that the system exhibits
quite different shapes depending on the initial orientations of $^{27}$Al that makes
subsequent dynamics so different. In the shrinking-$y$ case (a), the neck is rather thin,
and it ruptures shortly within about 1~zs ($t=1.6$--2.67~zs). Although the sticking time
is relatively short, more than 180$^\circ$ rotation is achieved in the reaction plane. In contrast,
in the shrinking-$z$ case (b), the elongated composite system exhibits a rather thick neck
structure ($t=1.27$~zs), which persists for much longer time (about 5~zs). From the second
row, each panel shows the composite system every after about 90$^\circ$ rotation in the
reaction plane ($t=2.13$--5.6~zs). From a careful look at the figure, we find that the system
has rotated as large as 720$^\circ$. It is interesting to observe that how the mass equilibration
is achieved in such a light asymmetric system. By looking at the density, we find that the
composite system shows rather complex shapes in the course of the collision. For instance,
one may follow the smaller subsystem that was initially the $^{16}$O nucleus, \textit{e.g.},
the top part of the dinuclear system at $t=0.6$~zs in Fig.~\ref{FIG:rho(t)}~(b). As time
evolves from $t=0.6$~zs to $t=2.67$~zs, the system rotates about $360^\circ$. However,
as seen in the figure ($t=2.67$~zs), the subsystem (top part) that was smaller at the initial stage
now looks larger than the other, indicating substantial nucleon exchanges from one nucleus
to the other in the dinuclear system (see Supplemental Material \cite{Supplement} for
movies of the reactions). Because of the nucleon exchanges and the long sticking time,
the system achieved the mass equilibration in the shrinking-$z$ case. The present results
indicate that deep-inelastic processes may substantially contribute to the generation of mass
symmetric fragments in this reaction, which may be associated with zeptoseconds sticking
time ($\gtrsim4$~zs, cf. Figs.~\ref{FIG:TKEL-theta} and \ref{FIG:Nave}) and perhaps
originate from shrinking-$z$-like collision geometries.

In the TDHF approach, it is not straightforward to evaluate angular distributions, due
to the semi-classical nature of the collective motions. Nevertheless, the calculation can
produce the total isotope production cross sections which is a very useful information for
understanding the reaction mechanism aspects. In the next section, we compare the
TDHF results with the experimental data for the integrated isotope production cross sections.

\section{CROSS SECTION COMPARISON}\label{sec:comparison}

As mentioned in Sec.~\ref{sec:expt}, we have deduced the $Q$-value- and
angle-integrated cross sections for various transfer channels from the measured
angular distributions (Fig.~\ref{o-n-ang}). In Fig.~\ref{FIG:sigmatot}, we present
the experimental data of the isotope production cross sections, which are represented
by red filled circles. The horizontal axis is the neutron number of the PLFs. In each panel,
the isotope distribution for a different proton-transfer channel ($\pm xp$) is shown. It is to
mention that, unlike in our earlier studies with heavy target \cite{bjroyPRC2015}, there
is no significant decrease in the cross section with the increase of the number of transferred
nucleons, indicating probably a different production mechanism for the present light
system at this incident energy.

In the TDHF approach, the usage of the particle-number projection method \cite{Projection}
allows us to evaluate production cross sections for primary reaction products before
secondary de-excitation processes from the TDHF wavefunctions \cite{KS_KY_MNT}.
The results of the TDHF calculations are shown in Fig.~\ref{FIG:sigmatot}~(a) by
histograms. Red, green, and blue lines, like before, correspond to the results associated
with the shrinking-$x$, $y$, and $z$ cases, respectively. Since a proper averaging over the
orientations requires a vast computational effort, here we show contributions from each initial
orientation examined. For comparison, results of a widely-used semiclassical model {\footnotesize
GRAZING} \cite{GRAZING} are also shown by gray shaded histograms, although the use of
{\footnotesize GRAZING} may not be suitable for collisions between such light nuclei and
at such high incident energy.

From Fig.~\ref{FIG:sigmatot}~(a), we find a significant discrepancy between the
measured cross sections and the TDHF results. Namely, TDHF substantially underestimates
the measured cross sections, especially, for reaction channels ($-xp$) ($x\ge2$). Moreover, the
experimental data indicate that the peak position of the cross sections should be shifted
toward the less neutron-number side for those processes. The discrepancies are very
similar to those observed for heavier systems \cite{KS_KY_MNT,KS_KY_Ni-U,Sekizawa(2017)}.
For the six-proton stripping reaction ($-6p$) calculated cross section is negligibly small
(below $10^{-5}$~mb) and can not be seen in the figure with the present scale settings.

As we have learned from the study of the TKEL spectra that the primary reaction products
are highly excited, secondary de-excitation processes via light-particle emissions may substantially
alter the final yields of the isotope distributions. A recently developed method, TDHF+GEMINI
\cite{Sekizawa(2017)}, enables the evaluation of production cross sections for secondary reaction
products. In the present work, the TDHF+GEMINI calculations have also been performed and
the results are shown in Fig.~\ref{FIG:sigmatot}~(b). In the calculations, average excitation
energy and angular momentum were utilized as detailed in Ref.~\cite{Sekizawa(2017)}.
We note that the ingredients of the statistical model have been parametrized and determined
so as to allow a good systematic description of the evaporation spectra for the entire mass region.
Detailed discussions on various modifications and fine-tuning of the model parameters that were
implemented in the {\footnotesize GEMINI}++ code can be found in Refs.~\cite{Charity2010,
Mancusi2010}. Nevertheless, one should keep in mind that the results may be still dependent on the
details of the model parameters. Since fine-tuning of those parameters is going beyond the scope
of this work, however, we utilize the default parameter setting for the {\footnotesize GEMINI}++
calculations.

From Fig.~\ref{FIG:sigmatot}~(b), we find that the inclusion of secondary de-excitation
processes significantly affects the isotope distributions. The orientation dependence is almost
washed out for the proton-stripping channels ($-xp$) after the inclusion of deexcitation effects.
The absolute value of the cross sections for the ($-1p$) and ($-2p$) reactions reasonably agrees
with the experimental data. For other channels, ($-3p$), ($-4p$), and ($-5p$), a significant
improvement of the agreement between the TDHF+GEMINI calculations and the experimental data
is observed and the overall trend, \textit{i.e.} the location of the peaks, nicely coincides with the
experimental data. However, the theory still underestimates the absolute value of measured cross
sections for these reactions as was observed also for heavier systems \cite{Sekizawa(2017)}.
We note that owing to the usage of the {\footnotesize GEMINI}++ de-excitation model, the dip
for $^8$Be production that decays into two $\alpha$ particles is correctly reproduced. In addition,
TDHF+GEMINI provides substantial cross sections also for $^4$He production (corresponding to the
$\alpha$ emissions) which can be seen as a prominent single peak in the ($-6p$;\;He) panel,
however, no experimental cross sections are available for comparison. We note that the
{\footnotesize GRAZING} results were obtained for limited reaction channels: it was unable
to provide cross sections in many of the proton-stripping channels ($-xp$), which indicates
that the model may not be suitable to apply to such a light system at energies substantially
above the Coulomb barrier.

To examine further, we have considered two types of division of excitation energy of the
fragments, (i) thermal division and (ii) equal division. For the case of the thermal division (i),
total excitation energy is shared as it is proportional to the fragment masses, while in the
equal division case (ii) the excitation energy is equally shared between two fragments.
The results presented in the Fig.~\ref{FIG:sigmatot}~(b) are actually correspond to the
case of the equal division. It was observed, by switching from the case (i) to (ii), an increase
in the magnitude of the production cross sections for ($-xp$) ($x\ge3$)
reaction channels as expected (the equal division gives higher excitation energy to the PLFs and thus
larger evaporation effects). As this gives a better agreement with the experimental
cross sections, some underestimation of evaporation effects within TDHF+GEMINI might be
the cause for observed discrepancy between the calculation and measurement, as discussed
in Ref.~\cite{Sekizawa(2017)}.

Lastly, it is important to note that the observed agreement between the TDHF+GEMINI
calculations and the experimental data are noteworthy, especially because no particular
effort was paid to reproduce the measurements. The present results indicate that the
deep-inelastic multinucleon transfer processes followed by light-particle emissions may
be the primary production mechanism for the proton-stripping processes in the
$^{16}$O+$^{27}$Al reaction at $E_{\rm lab}=134$~MeV.

\section{SUMMARY AND CONCLUSIONS}\label{sec:summary}

Deep-inelastic multinucleon transfer processes in the $^{16}$O+$^{27}$Al reaction
at $E_{\rm lab}=134$~MeV have been investigated both experimentally and theoretically.
The differential cross sections and TKEL distributions for various PLFs have been measured. The
TKEL up to about 70~MeV is observed, which indicates the deep-inelastic character of the reaction.
The measured angular distributions are forward-peaked and decrease exponentially as the scattering
angle increases. The observed behavior supports a picture of a dinuclear system formation and its
subsequent decays, similar to quasifission dynamics routinely observed in much heavier systems.

For a deeper understanding of the reaction dynamics, we have carried out a detailed theoretical
analysis based on the TDHF theory. From the results, we have found that the global features of
the reaction, such as TKEL and the scattering angle, do not depend much on orientations of the
deformed $^{27}$Al. On the other hand, the sticking time and the fragment masses show
noticeable orientation dependence. In an impact parameter region very close to the fusion
threshold, we have found orbiting-type dynamics, where the composite dinuclear system
rotates more than 360$^\circ$ in the reaction plane. Depending on initial orientations,
we observed generation of mass symmetric fragments after a relatively long sticking time
($\gtrsim4$~zs). This observation suggests that a significant contribution of deep-inelastic
collisions may be present in the symmetric fusion-fission-like reaction products at this incident energy.

The isotope production cross sections obtained from the measurement and those from the TDHF
calculations have been compared. By applying a recently developed method, TDHF+GEMINI
\cite{Sekizawa(2017)}, production cross sections for secondary products were evaluated.
From the comparison, we have found a reasonable agreement between the experimental
and theoretical cross sections, especially peak positions of the isotope distributions. The
agreement indicates that the deep-inelastic multinucleon transfer followed by light-particle
emissions is the primary reaction mechanism in the present case.

In conclusion, the present results suggest that even with a ``light+light" system at energies
well above the Coulomb barrier we can study a rich many-body reaction mechanism that nicely
mimics the one observed in heavy systems: \textit{e.g.} dinuclear system formation, multinucleon
exchanges, energy dissipation as well as mass equilibration processes. It would be promising to
continue this experimental program along with the TDHF analysis to further develop our
understanding of the underlying mechanism in low-energy heavy ion reactions.

\begin{acknowledgements}
The authors wish to thank the operation staff of the BARC-TIFR Pelletron-Linac facility, Mumbai, for the
excellent support during the experiment.  One of the authors (BJR) is thankful to Sangita Hazarika and Srimita Saha,
students from Gauhati University, for their interest and involvement in the initial stages of the data analysis.
One of the authors (KS) acknowledges support of Polish National Science Centre (NCN) Grant, decision
No.~DEC-2013/08/A/ST3/00708. This research work used computational resources of the HPCI system
(HITACHI SR16000/M1) provided by Information Initiative Center (IIC), Hokkaido University, through
the HPCI System Research Projects (Project IDs: hp160062 and hp170007).
\end{acknowledgements}

\newpage

\vspace{5mm}
\begin{center}
{\bf Supplemental Material for:}
{\bf ``Deep-inelastic multinucleon transfer processes in the $^{16}$O+$^{27}$Al reaction"}\\[-1mm]
\end{center}
\begin{small}
\noindent
In this Supplemental Material, we provide a list of supplemental movies of the $^{16}$O+$^{27}$Al
reaction at $E_{\rm lab}=134$~MeV obtained from the TDHF calculations. In the filename of each movie,
the impact parameter $b$ in fm and the initial orientation of $^{27}$Al (shrinking-$x$, $y$, or $z$) are
indicated. In the movies, 1 second of play time corresponds to 0.27 zeptoseconds in the simulation.
\end{small}

\vspace{6mm}
\noindent
\textcolor[rgb]{0.75,0,0}{\underline{\bf The shrinking-$\bm{x}$ case:}}
\begin{enumerate}
\setlength{\parskip}{1mm}
\setlength{\itemsep}{1mm}
\item{\tt 16O+27Al\_E134\_b8.000\_shrinking-x.mp4\\\url{https://youtu.be/rzmvv_bPE80}}
\item{\tt 16O+27Al\_E134\_b7.500\_shrinking-x.mp4\\\url{https://youtu.be/52AjwjX_iXk}}
\item{\tt 16O+27Al\_E134\_b7.025\_shrinking-x.mp4\\\url{https://youtu.be/-MaBOAgLEnE}}
\item{\tt 16O+27Al\_E134\_b7.024\_shrinking-x\_fus.mp4\\\url{https://youtu.be/D53LopW51BE}\\}
\end{enumerate}
\vspace{-3mm}
\textcolor[rgb]{0,0.50,0}{\underline{\bf The shrinking-$\bm{y}$ case:}}
\begin{enumerate}
\setlength{\parskip}{1mm}
\setlength{\itemsep}{1mm}
\item{\tt 16O+27Al\_E134\_b7.000\_shrinking-y.mp4\\\url{https://youtu.be/5w3ZZoH08rA}}
\item{\tt 16O+27Al\_E134\_b6.500\_shrinking-y.mp4\\\url{https://youtu.be/sXeoS6KNqIs}}
\item{\tt 16O+27Al\_E134\_b6.348\_shrinking-y.mp4\\\url{https://youtu.be/ZgPxRQtSX8w}}
\item{\tt 16O+27Al\_E134\_b6.347\_shrinking-y\_fus.mp4\\\url{https://youtu.be/CvnicMpZrJw}}
\end{enumerate}
\vspace{-1mm}
\textcolor[rgb]{0,0,0.80}{\underline{\bf The shrinking-$\bm{z}$ case:}}
\begin{enumerate}
\setlength{\parskip}{1mm}
\setlength{\itemsep}{1mm}
\item{\tt 16O+27Al\_E134\_b8.000\_shrinking-z.mp4\\\url{https://youtu.be/b5kXqdSxGDU}}
\item{\tt 16O+27Al\_E134\_b7.500\_shrinking-z.mp4\\\url{https://youtu.be/OLgC57c7cdM}}
\item{\tt 16O+27Al\_E134\_b7.000\_shrinking-z.mp4\\\url{https://youtu.be/riJVSs3XEiU}}
\item{\tt 16O+27Al\_E134\_b6.899\_shrinking-z.mp4\\\url{https://youtu.be/ks7ap1Be4dU}}
\item{\tt 16O+27Al\_E134\_b6.898\_shrinking-z\_fus.mp4\\\url{https://youtu.be/db3GJ-pi_Sc}}
\end{enumerate}


\end{document}